\documentstyle[ichep,12pt]{article}

\title{\normalsize THE NATURE OF THE CONTINUUM LIMIT \\
       IN STRONGLY COUPLED QUENCHED $QED$.
        \thanks{Partially supported by NSF-PHY 92-00148}}
\author{
 Maria-Paola Lombardo,
 John B. Kogut \\
 Department of Physics \\
 University of Illinois at Urbana-Champaign \\
 1110 West Green Street, Urbana, IL 61801, U.S.A.\\
 Aleksandar Koci\'{c}\\
 Theory Division \\
 C.E.R.N.\\
 CH-1211 Geneva 23, Switzerland \\
 K.C. Wang\\
 School of Physics \\
 University of New South Wales \\
 Kensington, NSW 2203, Australia }
\begin{document}
\finalcopy
\maketitle

\abstract{
   We review the results of large scale simulations\cite {KKLW}
 of noncompact
quenched $QED$ which use spectrum and Equation of State calculations
to determine the theory's phase diagram, critical indices, and
continuum limit. The resulting anomalous dimensions are in good
agreement with Schwinger-Dyson solutions of the ladder graphs of
conventional $QED$ and they satisfy the hyperscaling relations expected
of a relativistic renormalizable field theory. The spectroscopy
results satisfy the constraints of the Goldstone mechanism and
PCAC, and may be indicative of Technicolor versions of the Standard
Model which are strongly coupled at short distances.}

\vskip-1pc
\onehead{INTRODUCTION}
What is the simplest, most elementary yet physical example of anomalous
dimensions? Perhaps it is the relativistic hydrogen atom, a subject we
all learn long before field theory. In
the Dirac theory the ground state wave function behaves at short distances
like $r^{(\gamma-1)}$ where
$\gamma-1= (1-Z^2\alpha^2)^{1/2}-1=-Z^2\alpha^2/2+...$.
\break
\newpage
\vglue 2pt
\noindent
The
unscreened Coulomb attraction has made
 the wave function more singular
than predicted by the Schrodinger equation.
Considering scale transformations,
we see that the interaction has changed the scaling dimension
 of the electron
field and an anomalous dimension $\eta=\gamma-1$ has appeared. The anomalous
dimension vanishes in the Schrodinger description of the hydrogen atom because
the kinetic energy
$p^2/2m$ dominates the interaction $e^2/r$
at short distances.
The electron wave function
has ``canonical" or free field dimensions in this
case. However, in the Dirac equation the relativistic kinetic energy is $p$
which scales in the same fashion as the potential and the interaction can
not be neglected at short distances.
$QED$ with massless fermions can be solved in the ladder ``approximation" and
much of the same physics emerges in a relativistic
setting\cite{BLL}. The appearance
of anomalous dimensions which depend continuously on the bare charge is
confirmed. In fact, the physics is very rich, complete with chiral symmetry
breaking, a Goldstone mechanism, etc. Again, we are not dealing with a real
field theory but the ladder approximation is a step in that direction and
it is thought to be a good approximation to a class of Technicolor theories
\cite{PES}.

$QED$ in the ladder approximation also poses a challenge to lattice gauge
theory which should eventually solve strongly coupled $QED$ including fermion
vacuum polarization. Can lattice simulations of the quenched theory formulated
on the lattice and simulated by first principle methods yield chiral symmetry
breaking and anomalous scaling laws in agreement with the continuum
formulation?
In this report we shall answer this question in the affirmative, although
the lattice approach has not mapped out the full parameter space of couplings
available to the continuum formalism. We shall also see how lattice
spectroscopy
calculations can be used to find anomalous dimensions and uncover the
non-triviality of this not-so-simple model.
% One-column figure
% One-column figure
\begin{figure}
\vspace{23pc}
\vbox{\parindent=0pt\baselineskip=12pt\footnotesize Figure~1.
Chiral equation of state for $\beta_c = 0.257$, $\delta = 2.2$
and $\beta_{mag} = 0.833$}
\label{F1}
\end{figure}
\begin{figure}
\vspace{23pc}
\vbox{\parindent=0pt\baselineskip=12pt\footnotesize Figure~2.
Equation of state for the $\rho$ mass. $M_\rho/t^\nu$ is plotted
versus $m / (\beta_c - \beta)^{\delta\beta_{mag}}$
for $\nu = 0.67, \beta_{mag} = 0.84$
and $\delta = 2.2$}
\label{F5}
\end{figure}
\onehead{THE MODEL}
The lattice $QED$ action $S_{QED}$ in its non-compact formulation reads:
\begin{eqnarray}
S & = & S_S + S_{SB}
\end{eqnarray}
where
\begin{eqnarray}
S_S & = & 1/2e^2 \sum_{\mu,\nu,n} F^2_{\mu\nu}(n) + \nonumber  \\
& &  1/2 \sum_{\mu,n}
\bar\psi(n)\eta_\mu e^{i\theta_{\mu}(n)}\psi(n+\mu) \nonumber \\
& & + h.c. \\
S_{SB} & = & m\bar\psi\psi
\end{eqnarray}

$\theta_{\mu}$ are the gauge variables -- oriented, real
numbers ranging from $-\infty$ to $+\infty$--,
$F_{\mu\nu}$ is the
field strenght tensor,
$\psi$ are the fermionic fields,  $\eta$  the staggered phases.
$S$ is thus controlled by two bare parameters,
$\beta =   1/e^2$, $e$ being the electromagnetic coupling, and the
fermion mass $m$. By using staggered fermions
 $S_S$ has, for any value of the lattice spacing $a$,  a continuous
chiral symmetry  which
is spontaneously broken
at a finite (i.e. non zero) value of the coupling $\beta_c$.
The mass term
$S_{SB}$ is an explicit symmetry breaking term
which is required by  technical difficulties
connected with the chiral extrapolation.
We thus sample the critical region, at small, but non-zero $m$.
In this region ($\beta \to \beta_c, m \to 0$)
the fundamental tools are the Equation of State
(EOS) and the scaling laws : by exploiting them  we will be able
to get information about the singularities occurring at $\beta =
\beta_c$, $m_q = 0$  by working at finite values of $m$ and $\beta_c - \beta$.
This is similar to the study of
critical phenomena on finite systems: because of
the finite size  the system is
always  in the symmetric phase,
thus in both cases (finite size/finite mass) we deal with single phase
systems, and in both cases appropriate scaling laws tell us the
physics of  the phase  transition.

\onehead{THE EOS AND THE SCALING LAWS}

The response of the system in the critical region to an external symmetry
breaking field is expressed by universal functions
of reduced variables: this is the general fact which allows the
computation of the critical coupling, and related exponents.
In our case the symmetry breaking term is the mass term $m\bar\psi\psi$
in the lagrangian:  the response functions we consider are the chiral
condensate itself (the natural order parameter for the chiral symmetry),
and the meson masses.

\twohead{EOS for the chiral condensate}
The EOS for the chiral condensate,
in full analogy with a ferromagnetic system \cite {AMI}
(just replace $m$ with an
external magnetic field, and $\bar\psi\psi$ with the spontaneous
magnetization), reads:

\begin{equation}
m/<\bar\psi\psi>^\delta = f (t/<\bar\psi\psi>^{1/\beta_{mag}})
\end{equation}
(Here and in the following $t = \beta_c - \beta$.)
Its usage is demonstrated in Fig.1 where all our data are plotted and nicely
fall on a universal curve when $\delta = 2.2$, $\beta_{mag} = 0.833$
$\beta_c = 0.257$. The universal behaviour is
in principle unique to the correct parameters, but  the errors on the
data  broaden the choice. Thus  one has to
check the persistence of the (apparent) universal behaviour when
adjusting $\beta_{mag}, \delta, \beta_c$: the errors on these
indices  are determined by this procedure.
% Two-column figure
\begin{figure*}
\vspace{23pc}
\centerline{\footnotesize Figure~ 3.
$\bar\psi\psi^{x}$ versus $M_\rho$}
\end{figure*}

\twohead{EOS for the masses}

Once we assume that the critical behaviour of the system is controlled
by only one macroscopic (diverging) scale length,
the equation of state for the masses easily follows:
\begin{equation}
 M_a = t^\nu G_a (m/t^{\delta\beta_{mag}})
\end{equation}
$M_a $ is any meson mass (different of course from the Goldstone boson)
whose reciprocal $1/M_a$ is to be identified with the correlation length of
the system, times an irrelevant constant.
In complete analogy with the
EOS for the chiral condensate,  we can determine
$\beta_c$, $\delta$ and $\nu$
(Fig.2).

The check of the crucial hypothesis
that the critical behaviour is controlled by one divergent length scale
(apparently a hopeless task) is done a posteriori
by verifying the hyperscaling relations among the critical exponents
and indeed in this case proved to be  correct
(Table 1, to be discussed later).

\begin{figure}
\vspace{23pc}
\vbox{\parindent=0pt\baselineskip=12pt\footnotesize Figure~4.
log $M_\pi$ vs. log $m$ for $\beta = (0.245, 0.250, 0.255, 0.260, 0.265)$
(top to bottom). The straight line superimposed are fits to the relations
$M_\pi= A m^x$. In the strong coupling region $x$ is consistent with the
PCAC prediction of $0.5$.}
\label{F4}
\end{figure}

\begin{table*}[t]
\begin{center}
 \begin {tabular} { l l l l}
 \multicolumn{4}{l}{Table~1. Critical indices and hyperscaling relations} \\
\hline\hline
 Index & Result from the simulation & Result from HS  & MF\\
\noalign{\vspace{2pt}}
  \hline
 $\beta_{mag}$  &0.86(3)&0.86(6) &0.5\\
 $\gamma = \beta_{mag} (\delta - 1)$ &1.00(5)&   &1.0\\
 $\delta$ &2.2(1)&2.16&3.0\\
 $\eta$   &0.5(1)&0.5&0.0\\
 $\nu$    &0.675&0.68(3)&0.5\\
 $ - 4\nu + 2\beta_{mag} + \gamma$             &0.1(1)  &0.0&0.0\\
 $(2-\eta)\nu/\gamma$                    &1.1(1)  &1.0&1.0\\
 $(2\nu - \gamma)/(\nu\eta  )$           &1.1(1)  &1.0&1.0\\
 $(6 - \eta)/(2 + \eta) - \delta$&0.13(20)&0.0&0.0\\
\hline\hline
\end{tabular}
\end{center}
\end{table*}
\twohead{The scaling laws}
The EOS's for the chiral
condensate and for the masses can be, in some sense, combined to give the
following scaling law:
\begin{equation}
<\bar\psi\psi> = C_a M_a^{\beta_{mag}/\nu} = C_a M_a^{d/2 - 1+ \eta/2}
\end{equation}
whose most noticeable characteristic is  the
dependence on just {\it one} parameter, the anomalous dimension $\eta$.
So, eq.(6) offers the possibility of a simple test of
trivial vs. non-trivial behaviour: $\bar\psi\psi ^{x}$
($x =
\nu/\beta_{mag}$) plotted versus
 $m$ gives a straight line (including the origin)
for the correct value of $x$.
Recall  that in four dimensions $\eta = 0$ if the system is described
by mean field behaviour, which thus corresponds  to $x = 1$.
In Fig. 3 we experiment with different $x$ values:
the $x$ corresponding to the best fit (solid)
coincides within  errors with the ratio of $\beta$ and $\nu$ from independent
determinations (this method gives the most precise estimate of
$\nu = 0.68(3)$); the dotted line,
corresponding to the correct value of $\beta_{mag}$
and to the mean field value for $\nu$, demonstrates the sensitivity
of the method to the exponents choice; $x = 1$ (dashed),
hence $\eta = 0$, is clearly ruled out.
$\eta$ turns out to be $\simeq 0.5$, compatible with the result
given by the hyperscaling relation between $\eta$ and $\delta$.

\onehead{SPECTROSCOPY}
The numerical values of the fermion and meson masses contribute
to build a coherent scenario: the dynamical breaking of  chiral
symmetry should be associated with the appearance of a Goldstone boson,
the approach to the
continuum limit can be tested by looking at the scaling plots,
which in turn give information on the level ordering.

\twohead{The Goldstone character of the pion}

In the strongly coupled, symmetry broken region, we expect the
usual PCAC behaviour. The square of the pion mass should be linear
in the quark mass (plus second order corrections), eventually vanishing
in the chiral limit. The pion decay constant, on the contrary, should stay
finite in the same limit.
The pion behaviour is shown in Fig. 4 : the
relation $m^2_\pi \propto m_q$ is well verified in the strong coupling
region, while deviations are observed at weak coupling.
Figure 5 shows $f^2_\pi$.

\twohead{Level ordering}

The mass ratios can be plotted one against the other: near the continuum
limit the details of the lattice discretization are lost, and the resulting
plots (scaling plots) are $\beta$ independent. In a bit more formal fashion,
we can derive this property (widely exploited in lattice QCD studies)
by building up mass ratios from the EOS in Section 2 above. We sample
here the results for $\sigma/\rho$ vs. $\pi/\rho$ for $\beta = (0.245,
0.250, 0.255)$ : within large errors, all the data fall on the
same plot (Fig. 6).
\begin{figure}
\vspace{23pc}
\vbox{\parindent=0pt\baselineskip=12pt\footnotesize Figure~5.
$f^2_\pi$ as a function of $m$ for  $\beta$ = $(0.245, 0.250, 0.255, 0.260)$}
\label{F9}
\end{figure}
\begin{figure}
\vspace{23pc}
\vbox{\parindent=0pt\baselineskip=12pt \footnotesize Figure~ 6.
Scale invariant plot  for
$(\sigma/\rho)^2$ vs $(\pi/\rho)^2$. (Squares, diamonds, crosses)
are $\beta = (0.255, 0.250, 0.245)$}
\end{figure}

Moreover, we can get information on the level ordering in the chiral limit:
from Figure 6, and the analogous ones for the other ratios, we found
$0 = M_\pi < M_{\sigma} < 2M_f < M_\rho < M_{a1} $. As discussed at length
in \cite{KKL} $M_\sigma < 2M_f $ is a peculiarity of non-trivial theories.

\onehead{SUMMARY}
We summarize in Table 1 the critical indices and the relations among them
dictated by hyperscaling. In the first column we give the results
from the simulation, in the second column
the hyperscaling prediction assuming the
other indices as input for the single index entries, in the third column
the mean field prediction (d=4).
These results, together with the ones from spectroscopy,
support a picture of non-trivial critical
behaviour, which is inferred both from the hyperscaling, and
from the large anomalous
dimension $\eta$.


\begin{thebibliography}{99}
\bibitem {KKLW} A.~Koci\'{c}, J.~B.~Kogut, M.~--P.~Lombardo and K.~C.~Wang,
``Spectroscopy, Scaling and Critical Indices in Strongly Coupled
Quenched $QED$'', CERN-TH.6542/92 ILL-(TH)-92-\#12, (1992).
\bibitem {BLL} W.~A.~Bardeen, C.~N.~Leung and S.~T.~Love, ``Aspects of
dynamical
symmetry breaking  in gauge field theories'',
{\it Nucl.Phys.} B323, pp.~493--512, (1989).
\bibitem {PES} M.~E.~Peskin, ``Beyond the standard model'', these Proceedings.
\bibitem {AMI} D.~J.~Amit, {\it Field theory, the renormalization group, and
critical phenomena}, (McGraw-Hill, 1978).
\bibitem {KKL}
A.~Koci\'{c}, J.~B.~Kogut and M.~--P~.Lombardo,
`` Universal properties of chiral symmetry breaking'',
ILL-(TH)-92-\#18, CERN-TH.6630/92, (1992).

\end{thebibliography}
\end{document}